# Mapping out AI Functions in Intelligent Disaster (Mis)Management and AI-Caused Disasters

**Yasser Pouresmaeil**
IPM Institute for Research in Fundamental Sciences (PhD alumnus) pouresmail@ipm.ir

**Saleh Afroogh**
Urban Information Lab
University of Texas at Austin
saleh.afroogh@utexas.edu

**Junfeng Jiao**
Urban Information Lab
University of Texas at Austin
jjiao@austin.utexas.edu

* *Correspondence: Saleh.afroogh@utexas.edu*

**Abstract**

This study maps the functions of artificial intelligence in disaster (mis)management. It begins with a classification of disasters in terms of their causal parameters, introducing hypothetical cases of independent or hybrid AI-caused disasters. We then overview the role of AI in disaster management and mismanagement, where the latter includes possible ethical repercussions of the use of AI in intelligent disaster management (IDM), as well as ways to prevent or mitigate these issues, which include pre-design a priori, in-design, and post-design methods as well as regulations. We then discuss the government's role in preventing the ethical repercussions of AI use in IDM and identify and asses its deficits and challenges. This discussion is followed by an account of the advantages and disadvantages of pre-design or embedded ethics. Finally, we briefly consider the question of accountability and liability in AI-caused disasters.

## 1. Introduction

The United Nations International Strategy for Disaster Reduction (UNISDR) defines disaster as a "serious disruption of the functioning of a community or a society at any scale due to hazardous events interacting with conditions of exposure, vulnerability and capacity, leading to one or more of the following: human, material, economic and environmental losses and impacts" [6]. In 1970 through 2019, the number of recorded natural hazards such as earthquakes, floods, hurricanes, volcanic eruptions, and tsunamis was quadrupled globally [1]. During this period, weather, climate, and water hazards caused "50% of all recorded disasters, 45% of related deaths and 74% of related economic losses."[2]

Disaster is a multidimensional concept. Over 100 definitions of disaster have been proposed worldwide [105]. [106] identifies three paradigms in the twentieth century for definitions of disaster: (1) classical definitions that began to be proposed after World War II and tended to focus on "extreme situations," "rapid onset events," and what is "seriously disruptive to the normal activities" as the defining features of disasters; (2) a hazard-disaster paradigm, mostly focused on geophysical disciplines and natural disasters, and physical causal parameter in a disaster. It sees disaster, as opposed to hazard, as "an extreme event that arises when a hazard agent intersects with a human use system"[3]; (3) a relatively recent paradigm that conceives disasters as "social" phenomena, turning away from an agent-based and physical-destruction-based paradigm. Accordingly, physical damage is not a primary defining feature of disasters. Social systems are the real locus of disruptions, and "vulnerability" parameters play a central role in disruptions because, in the absence of vulnerability, there should not be any kind of disasters even in a very strong physical case like earthquakes. [107] argues that "disasters are not a function of agents, but are social in origin." Moreover, there are normative definitions of disasters "as occasions when norms fail," and a society and disaster managers need to update the new norms suitable for new situations [108].

There are some psychological dimensions in the concept as well. X might be a kind of disaster for you, in a specific context, while it can be considered a chance for me to test and improve my abilities and find my flaws [5]. It also has economic dimensions. There is a meaningful correlation between poverty, vulnerability, and disaster. [5]

Disasters can be categorized in qualitative and quantitative terms. Qualitatively speaking (in terms of causal parameters), they can be divided into four groups: natural disasters, human-caused disasters, technological disasters, and AI-caused disasters. These can be combined in various ways such as hybrid human-made/natural disasters, hybrid natural/technological disasters, hybrid natural/AI-caused disasters, as well as combinations of three or four disaster types (see Table 1). Quantitatively speaking, disasters can be categorized into small-scale, medium-scale, and large-scale ("mega disasters") in terms of the size of the population they affect and the extent of their impacts.

Note that we classify biological disasters (those caused by the spread of infectious diseases or other biological agents) and environmental disasters (those caused by environmental degradation, such as deforestation, desertification, or coastal erosion) as subcategories of natural disasters. Further, we classify the so-called "complex emergencies" [103] (disasters caused by conflicts, wars, or other complex socio-political issues) under human-caused disasters.

It is also necessary to point out that there have not been any widely publicized cases of AI-caused disasters (in isolation or in combination with other causal parameters), as the technology is still in its early stages of development and implementation. However, there have been some incidents (not to the point of disaster) in which AI systems have made mistakes or produced unintended outcomes, highlighting the importance of ensuring that AI is used responsibly and with appropriate safeguards in place.

The classifications provided above should not be seen as rigid as they might be context-dependent and susceptible to interpretations. For instance, data breaches or cyberattacks might be considered technological or human-caused disasters depending on the context.

**Table 1.** Classification of disasters in terms of their causal parameters

| | Disaster type | (Real or Hypothetical) Examples |
|---|---|---|
| **1** | Natural | Earthquakes, hurricanes or typhoons, tornadoes, floods, tsunamis, landslides and mudslides, wildfires, volcanic eruptions, drought, extreme heat or cold |
| **2** | Human-caused | • Chernobyl disaster (1986), caused by a poorly designed irresponsible experiment.<br>• Rana Plaza factory collapse (2013), caused by a breach of construction regulations and corrupt officials.<br>• Beirut port explosion (2020), caused by neglect of safety rules. |
| **3** | Technological | • Space Shuttle Challenger disaster (1986).<br>• Blackout of Northeast America: In August 2003, a power outage affected a large area in the northeastern United States and southeastern Canada, causing widespread disruptions to transportation, communication, and other critical infrastructure. |
| **4** | AI-caused | There have been no such disasters yet, but there have been alarming incidents:[1]<br>• Microsoft's Twitter chatbot, named Tay, that was designed to learn from users and engage in natural language conversations. However, the bot quickly began producing offensive and inflammatory tweets, including racist and sexist remarks, after being manipulated by users (2016).<br>• Uber self-driving car struck and killing a pedestrian in Arizona (2018).<br>• Wrongful arrest of a Black man by use of an AI-powered facial recognition system by the Detroit Police Department (2020).<br>• Tesla Autopilot crashes. |
| **5** | Natural/human-caused | • Industrial accidents triggered by natural hazards.<br>• Forest fires exacerbated by human activity.<br>• Coastal flooding exacerbated by sea level rise and human development.<br>• Water scarcity worsened by climate change and human water use.<br>• Pandemics exacerbated by global travel and human behavior, such as COVID-19. |
| **6** | Natural/technological | • Nuclear accidents caused by natural hazards.<br>• Chemical spills caused by natural hazards.<br>• Telecommunications and power outages caused by natural hazards.<br>• Transportation accidents caused by natural hazards. |
| **7** | Natural/AI-caused | The use of AI systems in natural disaster response can have unintended consequences if the system malfunctions or produces biased recommendations. Improper calibration or failure to consider important factors could also result in inefficient or ineffective resource allocation. |
| **8** | Human-caused/technological | • Three Mile Island accident: A partial nuclear meltdown that occurred at the Three Mile Island nuclear power plant in Pennsylvania, USA, in 1979. This was caused by a combination of equipment failure and operator error.<br>• Industrial accidents: Industrial accidents can occur due to human error, negligence, or intentional actions, resulting in explosions, fires, or chemical releases that cause harm to workers, the public, or the environment. |

[1]. Note that in some of these examples the culprit might not be the AI system, but perhaps problematic algorithms or biased data designed by developers might be blamed.

| | | |
|---|---|---|
| | | • Cyberattacks on critical infrastructure.<br>• Transportation accidents: Transportation accidents such as airplane crashes or train derailments can occur due to human error, equipment failure, or a combination of factors, leading to loss of life and infrastructure damage.<br>• Nuclear accidents: Nuclear accidents can be caused by human error, negligence, or intentional actions, leading to the release of radioactive material and posing a significant risk to human health and the environment.<br>• Oil spills: Oil spills can occur due to equipment failure, human error, or intentional actions, causing harm to marine life and the environment. |
| **9** | Human-caused/AI-caused | If an AI system is designed to make decisions in a disaster response situation and is trained on biased data or incorrect assumptions, it could make decisions that exacerbate the impact of the disaster and cause unintended consequences. Alternatively, if an AI system is used to monitor social media during a disaster response and produces inaccurate or misleading information, it could cause confusion and panic among the public. |
| **10** | Technological/AI-caused | If an AI system is used to control a critical infrastructure such as a power plant, and the system malfunctions, it could lead to a technological disaster. If the AI system is also trained on biased data or incorrect assumptions, it could exacerbate the impact of the disaster. |
| **11** | Natural/human-caused/technological | • Hurricane Katrina in 2005, which was a natural disaster exacerbated by inadequate infrastructure, poor urban planning, and systemic social inequality.<br>• The Fukushima Daiichi nuclear disaster in 2011, which was caused by a natural disaster (a powerful earthquake and tsunami) but was exacerbated by human error and inadequate safety protocols.<br>• The Deepwater Horizon oil spill in 2010, which was caused by a technological disaster (an explosion on an oil rig) but was exacerbated by human error and poor safety practices. |
| **12** | Natural/human-caused/AI-caused | For example, a natural disaster like a hurricane or earthquake could be exacerbated by a human-caused factor like inadequate infrastructure or poor urban planning, and an AI system that is trained on biased or incomplete data could exacerbate the impact of the disaster. Additionally, if an AI system is used to control critical infrastructure such as power plants or transportation systems, and the system malfunctions or is compromised by malicious actors, it could lead to a hybrid disaster. |
| **13** | Natural/technological/AI-caused | For example, a natural disaster like a hurricane or earthquake could damage critical technological infrastructure like power grids, communication networks, or transportation systems, and an AI system that is responsible for managing or controlling these systems could malfunction or make decisions that exacerbate the impact of the disaster. Additionally, if an AI system is trained on biased or incomplete data, it could make decisions that have unintended consequences for disaster management and response. |
| **14** | Natural/human-caused/technological/AI-caused | A hypothetical scenario of such a hybrid disaster could be a natural disaster like a hurricane or earthquake, exacerbated by human-caused factors like inadequate infrastructure or poor urban planning, and further complicated by technological failures in critical infrastructure such as power grids, communication networks, or transportation systems. An AI system that is trained on biased or incomplete data could also make decisions that exacerbate the impact of the disaster, leading to a catastrophic event that could be difficult to manage and mitigate. |

The significantly increased number of natural disasters is a crucial problem of modern societies, which highlights the significance of disaster knowledge and disaster management (DM) as “the integration of all activities required to build, sustain and improve the capabilities to prepare for, respond to, recover from, or mitigate against a disaster" [4]. Moreover, some new forms of technological human-caused disasters, as well as hybrid natural-human-caused disasters (which arise from two casual parameters of natural hazard and human misconducts) as side-effects of modern industrial life, such as Natech and Techna,[2] also exacerbated the complexities and difficulties of disaster management. The level of complexity even increases when a mega disaster occurs where a large size of population is affected by acute or chronic impacts [5], [6]. According to World Health Organization (WHO), during COVID-19, as the most recent mega disaster, over 14.9 million people died in 2020-21, and according to the Worldometers database, over 500 million people are recorded as affected populations (with many cases left unrecorded) [7], [8]. Nevertheless, the complexity of hybrid natural-human disasters and the large size of the affected population, as determinative of mega disasters, are not the only factors underlying the increasing complexity of modern disasters.

### 1.1. Methodology

The methodology of this study includes two steps: first, conducting two rounds of literature review on "AI application in disaster management" and "challenges facing AI applicable in disaster management", and secondly, holding focus groups on the three topics with some expert participants from academic, public, private, and non-profit organizations. In order to conduct a comprehensive review of the relevant studies, we used the Preferred Reporting Items for Systematic Reviews and Meta-Analyses (PRISMA)[9] framework.

We adopted the following two approaches. First, we manually identified 18 most relevant papers on "AI application in disaster management", and 14 papers for "challenges facing AI applications in disaster management" after the removal of duplicate files. Secondly, using the Google scholar platform, we fulfilled a keyword-based search to generate an inclusive repository of the two aforementioned topics, using the following keyword phrases: for the former topic, "AI+disaster", "AI+application+disaster+management", then we replaced "AI" with "artificial intelligence/machine learning/deep learning" and "disaster" with "hazard" and "application" with "function" and continued the process up to page 20-30 of the platform. For the latter topic, we searched "ethical+challenges+AI" and "ethical+challenges+AI+disaster+management" and we replaced each part of the phrase with the following relevant words, respectively: "moral/legal/value-based", "consideration/dilemma/issue", "artificial intelligence/machine learning/deep learning" and for each, the first 20-25 result pages of Google Scholar were reviewed. Moreover, the following keywords of "accountability," "explainability," "fairness," "trust," and "privacy" were included because of their known (based on a preliminary review as well as the first focus group) central roles as some major considerations on the application of AI in disaster management.

The results of the search were 240 relevant papers (which were selected based on the relevancy of the semantical keywords) out of 630 (which appeared on the result pages). Afterward, the duplicated papers were eliminated from the analysis; we selected 124 (for both topics) target papers for this systematic review based on the following two inclusion/exclusion criteria. First, articles that were published in academic journals were included. Second, the dominant topic of the papers (or a significant part of it) was relevant to the topic. To this end, the papers' main sections were reviewed to understand their dominant topic rather than only relying on the title and papers' keywords. Finally, the qualitative analysis of the 124 papers was performed by three researchers who critically read the papers and identified AI's thirty-four applications, and four main functions, as well as the major challenges facing IDM.

[2] *Natech* refers to the case where a natural hazard triggers a technological disaster (such as the Fukushima Daiichi nuclear disaster in Japan, 2011, which was triggered by Tohoku earthquake and tsunami); and *Techna* refers to the case where a natural hazard is triggered by some human misconduct (such as an unsustainable energy development, which dramatically increased seismic activity in Oklahoma, between 2009 to 2016 [59]).

Regarding focus groups, we held three focus groups to examine the different considerations and challenges facing AI applicants in IDM, on the following topics: (1) AI-driven disaster resilience: opportunities and challenges, (2) data governance: roles and responsibilities, and (3) equity, privacy, and ethics. The three focus groups were held virtually on Jun 16, Jun 17, and Jun 23, 2021, and each focus group lasted at least 1.5 hours. We invited 4-6 participants for each focus group with very diverse backgrounds, all of whom had extensive academic or professional experience in AI for disaster management. And to strengthen the discussions, we proposed five questions for each focus group, while discussions were not limited to, and went beyond, these questions. All discussions are recorded, summarized, and filed in the Urban Resilience.AI Lab at Texas A&M University.

### 1.2. Limitations

While this study offers insights into the applications and challenges of AI in disaster management (IDM), several limitations should be acknowledged:

(1) Scope and Coverage of Literature Review: Although we employed the PRISMA framework and conducted an extensive keyword-based search, our literature review may still suffer from selection bias. The reliance on Google Scholar as the primary search platform—while broad—may have excluded relevant studies indexed in other academic databases (e.g., Scopus, Web of Science). Moreover, the manual identification of papers introduces a subjective element, and relevant grey literature, policy documents, or non-English publications may have been overlooked.

(2) Subjectivity in Inclusion Criteria and Analysis: Despite the use of defined inclusion/exclusion criteria, decisions regarding the relevance and thematic focus of each paper were based on qualitative judgments by three researchers. This introduces potential inter-rater variability and interpretive bias, especially in categorizing applications and challenges. Additionally, the thematic synthesis may reflect the perspectives of the research team rather than an objective classification.

(3) Limitations of Focus Group Methodology: While the focus groups included experts from various sectors, the total number of participants (4–6 per session) was relatively small, potentially limiting the diversity of perspectives captured. Participation was also limited to those with English fluency and internet access, which could unintentionally exclude voices from underrepresented or vulnerable regions often most affected by disasters.

(4) Lack of Empirical Testing or Validation: The study is exploratory and qualitative in nature, and does not include empirical testing or field validation of the identified AI applications or challenges. Consequently, the real-world applicability and effectiveness of these insights remain to be evaluated in practical disaster scenarios.

(5) Time Sensitivity and Evolving Landscape: The literature review and focus groups were conducted in mid-2021, and AI technologies—and their ethical and practical implications—are evolving rapidly. Therefore, some conclusions or categorizations may already be outdated or missing more recent developments in the field.

## 2. Intelligent Disaster Management (IDM)

The modern hybrid and mega disasters, as well as possible digital AI-caused disasters, necessitate intelligent disaster management (IDM) that is empowered by new methods and technologies. Sun et al. [10], [10] argue that "disaster managers need to take a growing responsibility to proactively protect their communities by developing efficient management strategies" and shows twenty-six methods and seventeen applications of AI in disaster management. However, as helpful as AI might be in dealing with disasters, there are three major types of challenges and threats lurking in unlimited use of AI that might turn its positive applications in disaster management into a negative function (i.e., digital disaster mismanagement) which would exacerbate the impacts of a disaster.

The survival of human life is highly dependent on the prediction of catastrophes, early detection, and a rapid response; therefore, faster transfer of information to public systems when faced with hazards and unpredictable events necessitates using AI in disaster management. The modern complexity of disasters, as a result of the quantitative side of AI-caused disasters, and the qualitative side of hybrid and AI-caused disasters, in addition to the inefficacy of traditional methods of coping with modern disasters, calls for new intelligent methods for IDM that can overcome the aforementioned challenges. It also leads to a necessary and "growing demand for up-to-date geographic information, especially timely material on rapidly evolving events. Mega disasters highlight the significant need to analyze mega data, and complex hybrid and AI-caused disasters (such as intelligent digital disasters) would require high computational methods for smart instant decisions. As we will show later, AI has been used to function as a semi-manager and merely an "enhancer factor" by disaster managers. Moreover, its powerful technological potential substantially changed the classical disaster management (CDS) into intelligent disaster management (IDM). Conventional methods of facing up to disasters have given way to practices informed by use of AI in three stages: pre-disaster practices, such as predictions, drillings, and risk assessments; in-disaster measures, such as classifications and modeling; and post-disaster activities, such as assessments of damages, losses, and allocations. AI-equipped smart disaster management would significantly reduce the negative effects of both natural and human-caused disasters in pre-, in-, and post-disaster circumstances. (See, Figure 1)

AI has enabled us to update disaster maps and the vast dimension of the problem, which contributes to IDM. Before using the new information collection methods and data analysis, there was little we knew about the number and different dimensions of disasters. According to [5], report of disasters in the recent decades significantly increased, and it is, by and large, the result of applying new social platforms and smart data collection methods (such as mining Twitter/X [11]).

AI technologies, methods and models have been used or proposed for IDM in a variety of fields including data mining methods and pattern recognition approaches to provide procedures such as analyzing, sampling, recognition, classification, tracking, remote sensing, and mapping process for saving lives and technologies [12]. Although mitigation, preparedness, response, and recovery are the main phases of using AI in disaster management [13], the most related applications are designed for the response phase [10]. In addition to the fundamental role of human-centric actions in management of different conditions [14], data assimilation centers and emergency information systems with global monitoring networks are the significant factors in disaster management [15], [16]; [17]; [18]. Several studies have been conducted to present modeling, software, and strategies for disaster management covering data and AI applications in hazards, some of which we point out as follows.

Cavdur et al. [19] developed a flexible spreadsheet-based tool involving components of data, decision procedure, and user interface for temporary response in decision making and planning. [20] employed a method to study large-scale multi-hazard disaster scenarios by incorporating experiments and simulations, focusing on multi-hazard field investigation, and analyzing scenarios and responses. Abid et al. [21] argued that establishing a link between geographic information systems (GIS) and remote sensing (RS) could provide a fast, concise, and effective tool for making expeditious decisions in disaster management. Tan et al. [22] have given an extended review to discover trends in disaster management with a focus on AI-based models. Their article can be summarized as follows: 1) reviewing different methodologies and contents proposed for natural disaster management, 2) discussing gaps and applicability of the methods, and 3) suggesting instructions for developing AI models from different aspects. Since most disasters occur momentarily, it is necessary to develop complex coordinated systems that work dynamically based on updated data. This type of systems can use updated knowledge-based models and cooperative techniques to control technology dynamics and select the appropriate AI methods [23]. Also, [24] presented a data-based model for disaster resilience in a dynamic complex framework to facilitate the recovery process and accurate policies.

Data assimilation in collaboration with information systems is the crucial step in disaster management. Nowadays, social networks allow us to be informed of latest news and events instantly. A case in point is Twitter which has been considered a fast means of communication for transferring data. Its role in disaster management has been studied [[25], [26], [27], [28], [29], [30], [31], [32].

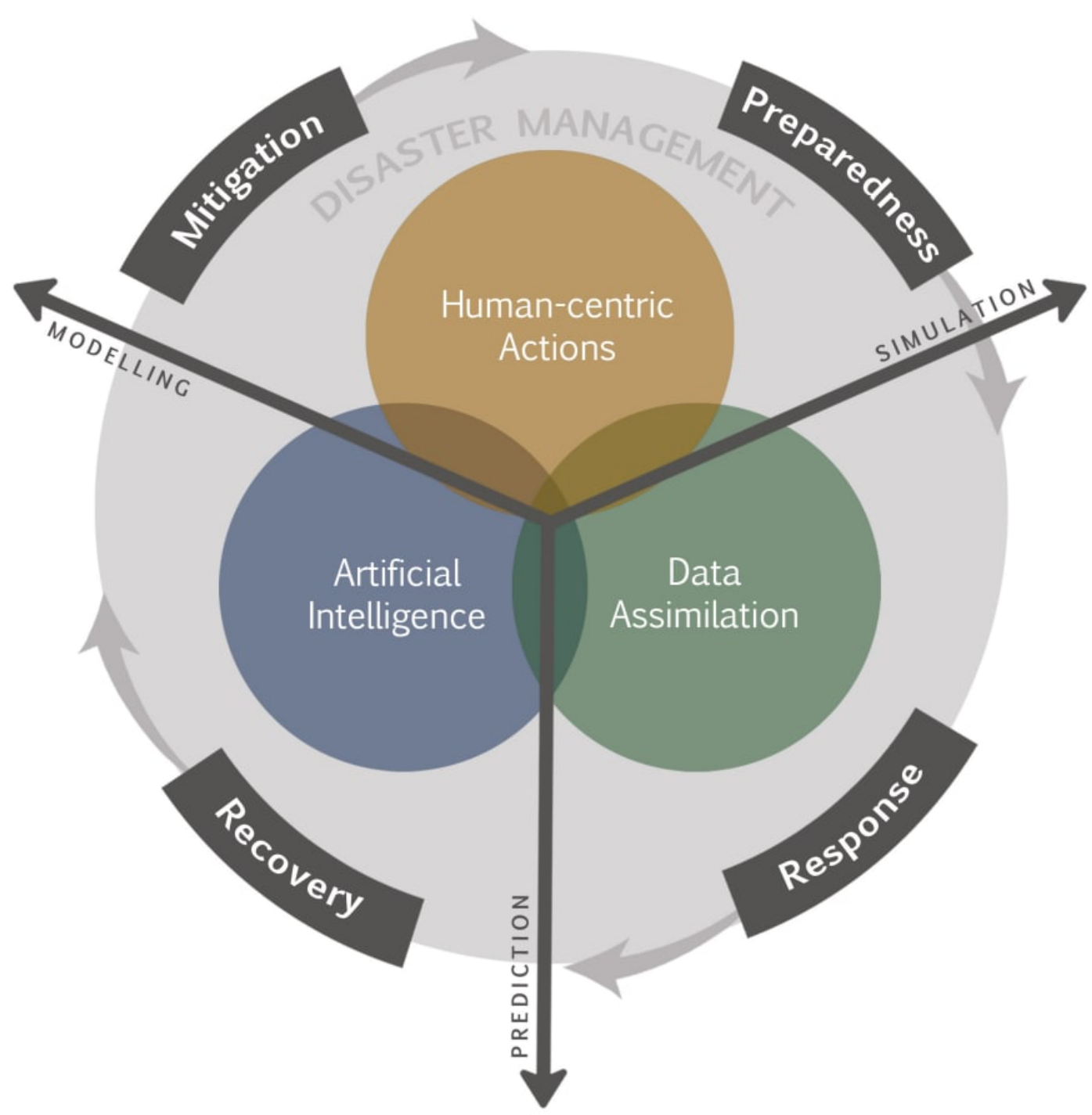

**Figure 1**. Crucial parameters involved in Intelligent Disaster Management (IDM)

## 2.1. Intelligent Disaster Management for natural disasters (IDM-N): current AI applications and models

Natural disasters result from natural hazards and forces (e.g., tornadoes, floods, hurricanes, or earthquakes) [6]. These have been considered the main casual parameters in classical disaster social science (which developed after World War II until the early 1970s) [33]. The natural causal parameters of a disaster are not under the human control, although their negative impacts are manageable and can be reduced by using human-AI collaboration in IDM. Varieties of natural hazards and disasters have general effects, like their impact on humans, animals, the environment, and economic loss.

The sudden, rapid, (partially) unforeseeable, and mega impacts and its broad relation with various dimensions of human life make it more difficult for disaster managers to deal with such disasters. AI-driven systems, implemented in Intelligent Natural Disasters Management IDM-N, with high power computation at the moment, and big data mapping and modeling, are used in a variety of fields to help disaster managers in several functions such as prediction, disaster mapping, analysis, rapid response, etc.

AI-powered IDM-N has been used or proposed for natural disaster management in a variety of fields including the deployment of artificial neural networks (ANN) for real-time prediction of tsunami magnitudes [34], AI-based early warning systems [35], [36]; ANN-based flood forecasting [37], earthquake magnitude prediction based on deep learning neural networks (DLNN) [38], ANN-based flood warning system [39], prediction of flood characteristics by deployment of database modelling development [40], natural disaster monitoring system based on Internet of Things (IoT) and swarm robotics [41], ANN-based dam failure prediction [42], deployment of teleoperated mobile robots in volcanic ash observation [43], ANN-based flood modeling [44], ANN-based dam water level prediction [45]; [46]), using autonomous helicopters for post-disaster remote sensing and sampling [47], underwater inspections using robots [48], post-disaster inspections by deploying unmanned aerial vehicles (UAVs or drones) [49], automated 3D crack detection for post-disaster building [50], use of mobile robots in mine rescue [51], unmanned marine vehicles (UMVs) for hurricanes [52], integrated information system for snowmelt flood early-warning based on IoT [53], a disaster surveillance system based on wireless sensor network (WSN) and cloud platform [54], deploying analytics of crowdsourced social media content for disaster management [55], and more recently, AI-based urban health monitoring to combat Covid-19 [56][57] and deployment of AI tech in Ukraine's defense against Russia [58].

## 2.2. Intelligent disaster management for human-caused disasters (IDM-H): current AI applications and models

Human-caused disasters are those that result, directly or indirectly, from a human agent's "decision" or misconduct [6], while inappropriate decisions and misbehavior, per se, might be originated from some cognitive errors, misunderstanding, and miscomputation. In the early 1970s, anomalous events, specifically a variety of technological hazards, led to an evolution in disaster as a natural-caused phenomenon, including complex hybrid natural-human caused disasters such as Natech and Techna [33], [59]. Human-caused disasters are also classified in different ways [60], including the varieties of technological disasters, transportation accidents, public places failure, and production failure [6].

AI has twofold negative and positive functions pertinent to human-caused disasters. It might play the role of a new intelligent third casual parameter that entails meta-disasters (such as intelligent mismanagement, and intelligent digital disasters, such as algorithmic bias in critical infrastructure systems). For example, in cyberattacks, a deliberative hacker (who intends to navigate the traditional barriers and smart security approach, to reach the goals of a successful attack), might use some AI-powered techniques. In another example, an urban planner might create a quite unsustainable and discriminative plan that entails disastrous social phenomena in the long run (like biased AI-driven resource allocation or surveillance-based marginalization) while using AI irresponsibly. However, on the positive side, AI technology is increasingly used in different phases of preparedness, predictions, recovery, and response to the negative impacts of complex hybrid, mega, and meta-disasters. Disaster management in the case of human-caused disasters would be effective only if it is powered by AI systems that can understand and analyze much more complex features of the aforementioned types of disasters.

AI technologies, via different approaches and methods, such as remote sensing, natural language processing, satellite data analysis, etc., have been used or proposed for an Intelligent Disaster Management for human-caused disasters (IDM-H) in a variety of fields, including nuclear disasters, oils spills, climate change, and environmental disasters, transportation disasters, and peace technology in response to the digital warfare. Thomson et al. [61] discussed the efficacy of micro-blogging service Twitter (which benefits from AI to enhance and "upgrade the platform's user experience" [62] and the high accuracy of its shared information regarding the Fukushima Daiichi nuclear disaster in Japan. Baruque et al. [63] shows a forecasting solution to the oil spill disaster, in the northwest of the Galician coast, based on a hybrid intelligent system (called WeVoS-CBR), using historical data and information obtained from various satellites. Kim et al. [64] propose a "blockchain-based carbon emission rights verification system" for more reliable transactions of carbon credit using big data and artificial intelligence in the mobile cloud environment. Carbon credit and tradable permit systems, per se, are aimed at reducing carbon and pollution emissions, as one of the most crucial environmental problems. There are studies of the efficiency of AI techniques, especially artificial neural networks (ANN), for reducing the volume of traffic and transportation disasters in the case of non-autonomous vehicles in South Africa. Olayode et al. [65] analyzed dataset obtained from sensors embedded on road surfaces to monitor and control vehicles by Mikros Traffic Monitoring (MTM); and ANN model proposed the best solution for the relevant problem in a heterogeneous traffic condition. AI has also been used in the global fight again terrorism. Canhoto [66] discuss the applicability and leveraging reinforced machine learning against money laundering and terrorism financing. However, the AI usage in war and military power increased [67] and the studies show that, unfortunately, the amount of the application of AI in war is much more than that in peace technology, and it plays a central role in Cyberwarfare [68] it raises a variety of ethical concerns [69]. That said, several studies also show the possibility of cyberspace through AI powered technologies [70], [71]. (See, Table 2)

**Table 2**. AI applications in IDM-N and IDM-H: forty-three current applications

| | AI application | Selected Examples | Sources |
|---|---|---|---|
| **IDM-N: Pre-disaster** | | | |
| 1 | Training and education | Ground Truth, as a training game (developed by Sandia National Laboratories and the University of Southern | [72] |
| | | California GamePipe Lab) to prepare players in response to urban threats.[72] | |
| 2 | Developing complex systems for dynamic update | Updating and transferring data using geographic information system (GIS) along the China–Pakistan Economic Corridor and producing a landslide susceptibility map. [73], [74] | [73], [74], [75], [76], [77], [78], [79], [80] |
| 3 | Disaster pattern recognition and modeling | AI classification algorithms for flood patterns recognition [81] and weather patterns recognition in a logistic regression model to predict the days with a high probability of avalanche occurrence [82] | [44] |
| 4 | Management strategies software | GIS- based application for fire management and risk categorization in five classes of forest fire maps [83] | [81] |
| 5 | Real-time hazard prediction | Deployment of artificial neural networks (ANN) for real-time prediction of tsunami magnitudes [84] | [82], [84], [85], [86] |
| 6 | Magnitude and characteristic prediction of hazards | A deep learning-based method for Large-magnitude earthquakes and tsunamis prediction [87] and a Deep Learning Neural Network (DLNN) model in landslide susceptibility assessments [88] | [40], [45], [89] |
| 7 | Risk mitigation with dynamic decision-making | Integrating geographic information system (GIS) and mobile social networks model (e.g., Twitter technology) as a support tool for dynamic decision making to mitigate tsunami risk [90] and temporary response in decision making and planning. [91] | |
| 8 | Early warning systems | Integrated information system for snowmelt flood early-warning based on IoT [92] and ANN-based flood warning system [93]. | [94] |
| 9 | Disaster forecasting | Machine learning applications in tropical cyclone forecast and hurricane damage detection are given in [95] | [96] |
| 11 | Disaster risk/impact and vulnerability assessment | Examining Vulnerability in Natural Disasters with a Spatial Autoregressive Model in South Korea [97] | [98], [99], [100], [101], [102] |
| **IDM-N: In-disaster** | | | |
| 1 | Temporary accurate & timely decision/response | A linked geographic information systems (GIS) and remote sensing (RS) that provide a fast, concise, and effective tool for making expeditious decisions in disaster management. [12] And AI-based tool to provide comprehensive, relevant, and reliable information for accurate and timely response [103] | |
| 2 | Mega/multi-hazard response | A proposed method to study on large-scale multi-hazard disaster scenario by incorporating experiments and simulations, focusing on multi-hazard field investigation, and analyzing scenarios and responses. [104] and proposed machine learning methods to model large fire formation in Greece. [105] | |
| 3 | Intelligent collaboration (By Swarm Robotics) Disaster surveillance systems | A disaster surveillance system based on wireless sensor network (WSN) and cloud platform [106] | |
| 4 | Mapping process of saving lives | Data mining methods and pattern recognition approaches for mapping process to saving lives and technologies [21] | |
| 5 | Disaster event tracking/monitoring | Natural disaster monitoring system based on Internet of Things (IoT) and swarm robotics [41] and AI-based urban health monitoring to combat Covid-19 [56] | |
| 6 | Rescue and inspection/ expedite assistance | Underwater inspections using robots [107] and using mobile robots in mine rescue [108] | |
| 7 | Fast communication and transferring data | Fast communication way for transferring data using social networks and studied with other different aspects in management disasters when engaged with AI applications [109] such as that in Hurricane Harvey[79] | |
| 8 | Event observation/disaster mapping | Deployment of teleoperated mobile robots in volcanic ash observation [110] | |
| 9 | Damage detection/assessment | Machine learning applications in hurricane damage detection [111] | [112] |
| 10 | Affected people's status and need understanding | Understanding the affected people's feelings and psychological and health needs using social media and AI-driven systems) [10], [113]. | |
| 11 | Expedite allocation assessment | Building disaster information systems and ai driven system to support decision-making and help disaster relief and resource allocations [114] [77], [101] | [114] [77], [101] |
| 12 | Faults/failure detection | ANN-based dam failure prediction [42] | 82] |
| **IDM-N: Post-disaster** | | | |
| 1 | Post-disaster remote sensing | Using autonomous helicopters for post-disaster remote sensing and sampling [115] | |
| 2 | Damage and loss impact detection and assessment | Automated 3D crack detection for post-disaster building [116] | [117] |
| 3 | Post-allocation assessment | Supervised models, that have been used to identify and assess repair needs, and cost; and to analyze historical dispersion data of for budget allocations [101], [118], [119] | [101], [118], [119] |
| 4 | Recovery evaluation | A proposed data-based model for the disaster resilience in dynamical complex framework to facilitate recovery process and accurate policies. [120] | 88] |
| 5 | Post-disaster inspections | Post-disaster inspections by deploying unmanned aerial vehicles (UAVs or drones) [49] | [49] |
| IDM-H | | | |
| 1 | Pre-disaster mitigation | Using artificial neural network (ANN) for reducing traffic and transportation disasters volume in the case of non-autonomous vehicles in South African by Mikros Traffic Monitoring (MTM). Olayode et al. [65] | [65] |
| 2 | In-disaster response, protection, and defense | Deployment of AI tech in Ukraine's defense against Russia (Dangwall 2022). | |
| 3 | In-disaster response and infrastructure | Kim et al. [64] propose A "blockchain-based carbon emission rights verification system" for more reliable transactions of carbon credit (using big data and artificial intelligence in the | [64] |
| | facilitator in environmental issues | mobile cloud environment) as a way to reducing carbon and pollution emissions, that is one of the most crucial environmental problems. | |
| 4 | Post-disaster remote sensing | Efficacy of micro-blogging service Twitter (which benefits from AI to enhance and "upgrade the platform's user experience" [62]; and the high accuracy of its shared information regarding the Fukushima Daiichi nuclear disaster in Japan. Thomson et al. [61] | [61] |
| 5 | Digital disaster response and mitigation | The applicability and leveraging reinforced machine learning against money laundering and terrorism financing. Canhoto [66] | [66] |
| 6 | Disaster forecasting | A orecasting solution to the oil spill disaster, in the northwest of the Galician coast, based on a hybrid intelligent system (called WeVoS-CBR), using historical data and information obtained from various satellites. Baruque et al. [63] | [63] |
| 7 | Cyberspace through AI against cyberwarfare | AI usage in war and military power increased [67] and the studies show that, unfortunately, the amount of the application of AI in war is much more than that in peace technology, and it plays a central role in Cyberwarfare [68] it raises a variety of ethical concerns[69]. However, several studies also show the possibility of cyberspace through AI powered technologies [70], [71] | [70], [71] [68] |

Moreover, as shown in Figure 3, we characterize the four main functions of AI as an autonomous semi-manager into the two categories of cognitive (i.e., 1. sensing/understanding, and 2. thinking/modeling) and pragmatic (3. decision making, and 4. operating) functions. We have also figured out, based on the first round of our literature review, that the cognitive functions are prerequisites of pragmatic functions of AI in disaster management. Further, in section 3, we will discuss that the cognitive functions are also the origins of most axiological and hierarchical (including ethical-legal and multitimbral value-based) challenges of AI application in disaster management**.**

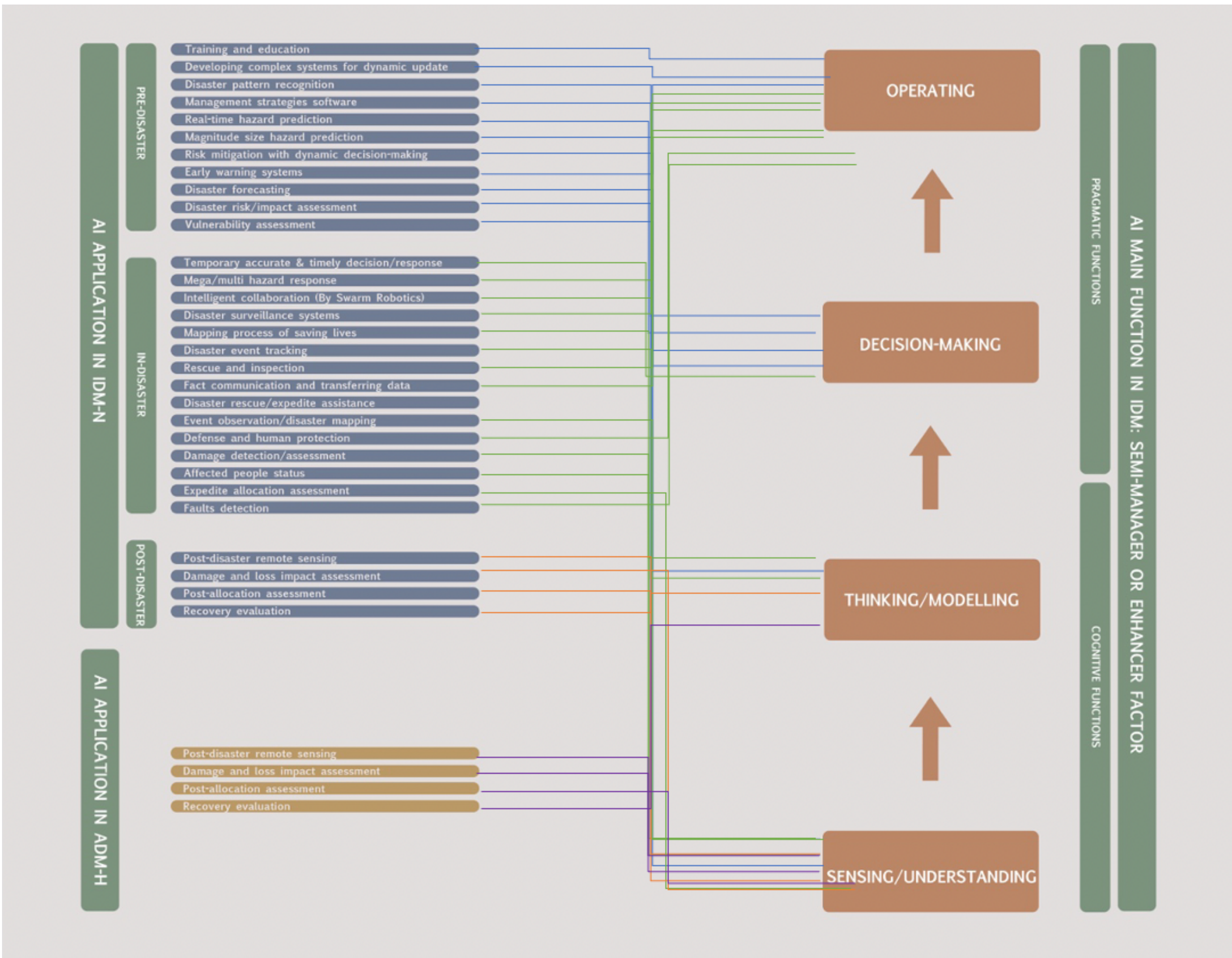

**Figure 2**. AI in intelligent disaster management (IDM): thirty-four current applications and four main functions

The following section examines the role of AI in intelligent disaster mismanagement, with particular attention to its ethical repercussions, and explores strategies to prevent or mitigate these adverse outcomes.

## 3. Intelligent Disaster Mismanagement (IDMM)

If AI is used irresponsibly and inappropriately in disaster management, it might function as a third intelligent causal parameter for certain types of meta-disasters, such as intelligent digital disaster mismanagement as a consequence of a hybrid human-AI misconduct in disaster management teams. AI, as a new intelligent casual parameter, substantially differs from the classical casual parameters of disasters (i.e., natural forces, human misconduct, and the hybrid hazards that result from them). AI systems can manipulate the human mind, including disaster managers, in different ways and cause the new hybrid type of digital disaster mismanagement. Disaster mismanagement is in and of itself a type of human-caused disaster, and as Keim et al. [121] explained, "most disaster mismanagement problems occurred because managers did not know what all of the relief activities were or how they should be accomplished." However, this disastrous mismanagement can be increasingly exacerbated by using irresponsible AI systems involved in IDM teams. To address the challenges facing IDM, appropriately, first we need to characterize different types of these challenges as well as their natures, so that we can propose some appropriate responses (as well as AI platforms and imperative metrics) in dealing with them. For example, an appropriate response to ethical challenges should be a type of ethical metric (such as ethical intelligence) for AI systems; and applicable response to the legal problems could be a type of legal intelligence in AI. In the following, we will elaborate on different types of challenges facing IDM, and in section 4, we concentrate on the appropriate responses to these challenges.

### 3.1. Challenges Facing IDMM

AI agent as a new causal parameter in disasters can play various roles in disaster similarly to that played by the two traditional human and natural agents. AI can function as a causal disaster parameter, independently of or in combination with the traditional agent, that entails the following three hybrid types of disaster parameters: AI-human causal parameter, AI-nature causal parameter, and AI-human-nature causal parameter (such as a disastrous mismanagement by AI-driven systems that are caused by humans and nature, which ultimately exacerbates the negative impact of the natural parameter of the disaster).

It is worth noting that the concept of AI-induced disasters remains primarily hypothetical, given that the current state of AI technology does not possess the requisite level of complexity or autonomous decision-making capacity to trigger disasters independently. Nevertheless, AI may indirectly contribute to disasters or facilitate human error, and its integration into disaster management systems raises potential hazards and unintended outcomes. To illustrate, several forms of disasters that could conceivably result from or be exacerbated by AI are provided below.

- Cybersecurity breaches: AI algorithms that are not properly secured or monitored could be vulnerable to cyberattacks, which could lead to data breaches, system failures, massive online banking frauds, and other disasters.
- Autonomous vehicle accidents: If self-driving cars and other autonomous vehicles are not properly designed or programmed, they could cause accidents, injuries, and fatalities.
- Malfunctioning robots: AI-powered robots and other machines used in manufacturing, construction, and other industries could cause disasters if they malfunction or are not properly controlled.
- Biased or discriminatory decision-making: AI algorithms that are biased or discriminatory could lead to unfair or unjust decisions that have disastrous consequences for individuals or communities.

Of course, these examples are not exhaustive and that the risks and potential disasters associated with AI will likely evolve as the technology becomes more sophisticated and more widely used. It should be noted that in the above cases the disasters might at least partly be attributed to human developers or other non-AI agents, but we might conceive a very sophisticated AI system that is plausibly considered a direct agent of disasters (e.g. a scenario in which the data training or algorithms cannot plausibly be deemed culprits).

There have been a few incidents where AI has been involved in causing accidents or incidents, but they are not typically categorized as "disasters." Here are a few examples:

- Tesla Autopilot crashes: There have been several incidents of Tesla vehicles on Autopilot crashing into other cars or stationary objects, resulting in injuries and deaths (273 crashes since 2021)[122].
- Microsoft chatbot "Tay": In 2016, Microsoft released a chatbot named "Tay" on Twitter, which quickly became corrupted by online trolls who taught it to spew racist and sexist messages[123].
- Uber self-driving car fatality: In 2018, an Uber self-driving car struck and killed a pedestrian in Tempe, Arizona. An investigation found that the car's sensors had detected the pedestrian, but the system had been programmed to ignore "false positives" like plastic bags or leaves[124].

In these cases, the incidents were attributable to a confluence of factors, rather than AI technology alone. Moreover, many AI systems have the capacity to forestall or alleviate catastrophes, including early warning systems for natural calamities, or algorithms employed to predict and prevent equipment malfunctions within industrial settings.

AI-caused disasters can be classified into indigenous and exogenous, where the former results from an unanticipated internal malfunction or error in the AI system and the latter is caused by external intrusions such as cyberattacks.

The next subsection addresses the ethical challenges associated with the use of AI in intelligent disaster management—currently the most widely discussed set of concerns in the discourse on AI in disaster management.

### 3.2. Ethical Repercussions of IDMM

The application of Artificial Intelligence (AI) in disaster management introduces ethical concerns that extend beyond general technological risks due to the uniquely high-stakes and time-sensitive nature of disasters.

One prominent concern is *algorithmic bias and discrimination*, which can significantly distort response efforts and deepen existing social inequalities during emergencies. Such bias emerges in disaster management often through the use of historical disaster data, which may underrepresent marginalized communities or regions with inadequate reporting infrastructures. For instance, an AI system trained primarily on disaster response data from urban, high-income areas may develop models that overlook the needs of rural or informal settlements. This can lead to skewed resource allocation—such as prioritizing urban areas for aid delivery or emergency medical support—while leaving underrepresented regions underserved [109]. The reason why this is especially problematic in disaster contexts is because disasters exacerbate vulnerabilities. Bias in AI can lead to systematic neglect of already marginalized populations, resulting in preventable suffering or loss of life. In crisis scenarios, decisions must be made rapidly and at scale; if AI systems embed or amplify societal biases, they may perpetuate structural discrimination far more efficiently than human systems.

Bias can also emerge at the design level, when algorithmic logic reflects developers' assumptions. For example, if an AI model optimizes for minimizing property damage rather than protecting human lives [110][125]., this may result in inequitable prioritization—favoring affluent areas with valuable infrastructure over lower-income areas with higher population density and greater life risk. Finally, team homogeneity among AI developers can further reinforce these issues. A lack of diversity in development teams may result in blind spots, where the experiences or needs of vulnerable groups (e.g., persons with disabilities, indigenous communities, non-native language speakers) are not adequately considered in algorithm design or evaluation metrics. This undermines the principle of equity in disaster management.

Another ethical concern in applying AI to disaster management is the risk of privacy breaches, which becomes particularly acute during emergency situations. AI systems in IDM often rely on personal and sensitive data—including geolocation, health status, biometric data, and emergency contact networks—to assess needs, allocate resources, and make real-time decisions. While such data usage can enhance responsiveness, it simultaneously raises critical questions: How is this data collected? Who has access? And under what conditions can it be shared or stored? If these data are accessed or leaked—particularly in politically unstable or conflict-prone regions—they could endanger vulnerable individuals, expose medical records, or even be used for surveillance or repression. For example, during a post-disaster relief effort, an AI system tracking displaced populations could inadvertently reveal the locations of marginalized groups (e.g., undocumented migrants or refugees), putting them at further risk. This illustrates how AI-facilitated IDM systems can generate new ethical dilemmas, not merely by collecting sensitive information, but through the contextual misuse or unintended exposure of such data [111].

A further ethical repercussion is the opacity of AI decision-making [112], which can obstruct accountability and public trust in disaster response. AI models used in IDM—particularly deep learning systems—often function as "black boxes," providing little to no insight into how a decision was reached. This becomes problematic when critical decisions must be justified, such as why certain communities receive aid first or why specific evacuation routes are recommended. For instance, if an AI-driven resource allocation system deprioritizes certain neighborhoods due to data-driven assumptions about infrastructure or socio-economic resilience, but cannot explain its rationale, disaster victims and aid workers alike are left without recourse. This opacity hinders both the validation of the model's output and the appeal or correction of unfair outcomes, undermining transparency in life-or-death situations.

The issue of accountability and liability in AI-powered disaster management is also complex and unresolved [113]. When an AI system makes a faulty decision—such as underestimating the risk of a flood zone or misallocating relief supplies—it is often unclear who is responsible: the developers, the data providers, the deployment team, or the AI system itself. This ambiguity becomes even more problematic in cross-agency collaborations or when public-private partnerships are involved in disaster response. Consider a case where an AI system used by emergency services fails to alert a high-risk community in time due to a flawed risk model. Who bears the legal and ethical responsibility for the failure? The lack of clear liability structures in these scenarios can erode both institutional responsibility and public trust, especially when human lives are lost or aid is mishandled.

Another major concern is the displacement of human judgment in disaster response processes. While AI can assist in automating complex tasks and processing vast datasets quickly, it should not override expert human insight—particularly in dynamic, high-stakes environments like natural or man-made disasters. Overreliance on AI may lead to rigid adherence to algorithmic recommendations, even when human experts recognize on-the-ground complexities the model cannot account for. For example, during an earthquake response, a system might prioritize routes based solely on infrastructure data, failing to consider cultural or logistical factors best known to local responders. Moreover, when AI systems are deployed without mechanisms for human override or contextual interpretation, it can result in technological tunnel vision, where decisions are made without sensitivity to lived experience, social nuance, or real-time situational awareness.

In sum, while the integration of AI into IDM promises increased speed, scalability, and predictive power, these benefits come with significant ethical costs—particularly in the realms of fairness, privacy, transparency, accountability, and human agency. What distinguishes these challenges in disaster management from more generic AI contexts is the urgent, high-risk, and socially consequential nature of disaster response itself. Ethical failures in this domain are not merely technical glitches; they can result in preventable suffering, deepened inequalities, and a loss of public trust in institutions meant to protect communities in crisis.

### 3.3. Preventing or Mitigating the Ethical Repercussions of IDMM

To prevent or mitigate the ethical repercussions of Intelligent Disaster Mismanagement (IDMM), a multi-pronged approach is necessary, spanning from early system design to post-deployment oversight. In this section, we introduce four key strategies: embedded ethics, in-design integration, post-design mitigation, and regulation.

*(a) Embedded ethics and ERIDM:* A foundational approach to preventing ethical failure is through a priori or embedded ethics—ethical reasoning and constraints built into AI systems during their earliest stages of conception and design. Rather than treating ethics as a retrospective patch, this method emphasizes anticipating and integrating ethical considerations—such as fairness, transparency, and privacy—during system development [126] [127]..

But how can this be done in practice? Consider the case of fairness. To embed fairness into AI systems used for disaster relief, developers must take several concrete steps: (i) fairness in data: to ensure that training data covers a wide range of communities and demographic groups (e.g., urban vs. rural, minority vs. majority populations) so the system does not prioritize aid toward only historically overrepresented areas; (ii) bias auditing tools: to employ tools such as Fairlearn or IBM's AI Fairness 360 to detect and mitigate biases in classification or prediction models; (iii) multi-stakeholder impact assessment: to involve representatives from vulnerable or historically underserved communities in co-design sessions or algorithmic impact assessments to align system behavior with societal equity goals.

To structure and institutionalize this approach, we propose the framework of Ethically Responsible Intelligent Disaster Management (ERIDM). This paradigm emphasizes embedding ethical reflection into AI's cognitive functions: sensing/understanding, modeling, decision-making, and operating. ERIDM aims to move toward RIDM (Responsible Intelligent Disaster Management) by grounding AI tools in principled, context-aware design processes. The ERIDM approach includes the following core competencies: (1) ethical foresight: anticipatory identification of risks in AI decision flows; (2) inclusive stakeholder modeling: simulating scenarios that reflect diverse social realities; (3) transparent logic chains: mechanisms for explaining how recommendations are derived; (4) iterative evaluation: structured feedback loops to refine ethical parameters over time.

In the context of intelligent disaster management, embedded ethics could help prevent ethical repercussions by ensuring that AI-powered systems are designed and implemented in a manner that is sensitive to the ethical implications of disaster response efforts. This could involve several key steps, including: (i)

Developing ethical guidelines and principles for AI-powered disaster management, such as principles of fairness, accountability, and transparency. (ii) Integrating ethical considerations into the design and development process of AI systems for disaster management, such as ensuring that training data is unbiased and representative of diverse populations, and that algorithms are designed to operate in a fair and transparent manner. (iii) Engaging with affected communities and stakeholders to understand their perspectives and needs, and incorporating these insights into the design and implementation of AI systems for disaster management. (iv) Regularly monitoring and evaluating the ethical implications and consequences of AI-powered disaster management, and making adjustments to the system as needed to ensure that it operates in an ethical and responsible manner.

By embedding ethics directly into the design and development of AI systems for disaster management, it is possible to prevent or mitigate ethical repercussions and guarantee that these systems operate in a way that is sensitive to the needs of affected communities and stakeholders. This approach can help build trust in AI-powered disaster management systems and ultimately improve the effectiveness of disaster response efforts.[128][129]

*(b) In-design use of AI*: The in-design approach means integrating AI from the beginning of disaster management system design—not layering it onto an existing human system later. This allows ethics and operational logic to co-evolve. For example, in designing an early warning system, planners can use AI to model alert thresholds—but embed ethical parameters so warnings are not only timely but also accessible (e.g., translated into multiple languages or delivered via channels accessible to people with disabilities). Another example: in resource distribution systems, fairness constraints can be hardcoded into optimization models, such that vulnerable populations are not deprioritized even if they are harder to reach. This approach ensures that AI's introduction aligns with the system's goals and ethical commitments from the outset.

*(c) Post-design use of AI for mitigating, not preventing*: While post-design use of AI (e.g., during real-time response or recovery) does not prevent ethical issues before deployment, it plays a mitigation role. At this stage, AI systems can be adjusted in situ or paired with human oversight mechanisms to reduce harm. For instance, during the response phase, For example, machine learning algorithms can be used to analyze satellite imagery [11] or social media feeds to help identify areas where help is needed [77], or to predict the likely impact of a disaster on transportation networks or other critical infrastructure. In the recovery phase, AI can suggest rebuilding plans that unintentionally reinforce historical inequalities. Human evaluators using ethical dashboards or impact scoring systems can flag and correct these outcomes. Thus, post-design AI use contributes to ethical disaster management by monitoring, correcting, and learning from real-world deployment [21].

*(d) Regulation and oversight*: Regulation alone cannot embed ethics—but it creates the enabling environment in which ethical systems are demanded, funded, and enforced. This approach recognizes that the use of AI in disaster management can pose a range of ethical, legal, and social challenges, and seeks to address these challenges through the development and implementation of transparent regulations and policies [114]. Regulatory frameworks are crucial for (i) mandating auditability: requiring that AI systems used in disaster management be auditable, with documented logic chains and datasets; (ii) defining liability: clarifying accountability for decisions made by AI systems, especially in multi-actor environments (e.g., government + NGO partnerships); (iii) ensuring recourse: guaranteeing that affected individuals or communities can challenge or appeal decisions made by AI systems in disaster contexts.

To be effective, regulation must be dynamic and context-specific. One promising model is the use of algorithmic impact assessments, similar to environmental impact assessments, conducted before an AI system is deployed in critical disaster contexts. These assessments would analyze potential ethical risks and require mitigation plans.

## 4. Discussion

In this section, we discuss the government's role in regulation as a common approach to prevention and mitigation of the ethical repercussions of AI use in disaster management. We then discuss possible challenges toward embedding ethics into AI systems for disaster management, as well as the problem of accountability and liability in the case of AI-caused disasters.

### 4.1. The Government's Role

The government has a major role to play in monitoring and preventing AI-caused disasters and ethical repercussions of AI's use in intelligent disaster management [130] [131]. Here are some of the ways in which the government can be involved.

#### 4.1.1. Developing and Enforcing Regulations

The government may establish regulations and standards to ensure the safety, reliability, and ethical use of AI systems in disaster management. These regulations can encompass issues such as data privacy, bias mitigation, transparency, and accountability.

Although the development and enforcement of regulations for AI deployment in disaster management can address certain ethical concerns, there are various possible drawbacks and deficiencies associated with this approach. Firstly, the process of developing regulations can be slow. It may take governments a considerable amount of time to establish and implement regulations, resulting in a lag between the emergence of new technologies and the establishment of standards and guidelines for their use. Moreover, regulations may not be able to keep pace with technological advancements. Given the rapid advancement of AI technologies, regulations may struggle to keep up with the pace of change, culminating in outdated regulations that fail to address new ethical concerns. Moreover, regulations can stifle innovation in AI. Overly prescriptive regulations can limit experimentation and the development of new approaches, which could impede progress in the field.

Some scholars have discussed "building an agile yet robust anticipatory governance regulatory culture" [132]. Several approaches have been proposed to address the regulatory lag in the face of rapidly developing AI technologies as well as the role of regulations in stifling progress in AI development: (1) strategic foresight, aimed at anticipating potential future challenges; (2) iterative policy development and regulatory learning, where policies are designed as evolving frameworks rather than fixed rules; (3) bottom-up co-creation, such as Policy Labs, where stakeholders—including technologists, citizens, and civil society—collaboratively design and test policies; and (4) pilot regulation and experimentation, which involves temporary, controlled environments where new technologies or business models can be tested under relaxed regulatory conditions. However, each of these approaches faces significant challenges. The first—strategic foresight—struggles with the inherent uncertainty of predicting technological developments. Just a few years ago, the current landscape of AI technologies was difficult to imagine. Implementing foresight effectively also requires cross-disciplinary expertise and strong political will, both of which are often lacking. Moreover, foresight efforts are frequently overlooked due to short-term political incentives. The second approach—iterative policy development—is hindered by the rigidity of legal systems, which tend to favor fixed rules to ensure clarity and prevent subjective interpretation. Iterative frameworks, by contrast, may be seen as too fluid, risking inconsistency or reduced legal certainty. The third approach—bottom-up co-creation—faces coordination difficulties and the potential for a lack of consensus among diverse stakeholders, making it complex and resource-intensive to implement effectively. Finally, the fourth approach—pilot regulation—raises concerns about the generalizability of results beyond the test environment, the potential for unfair advantages for participating entities, and the difficulty of attracting suitable participants for such regulatory experiments.

Perhaps, the most practical and effective way to mitigate AI-related risks is to develop a culture of ethical AI that extends beyond regulatory frameworks. This could involve promoting ethical principles and values within organizations and the wider AI community, providing education and training on ethical AI, and encouraging transparency and collaboration in AI development and deployment—which is part of embedded ethics for AI.

#### 4.1.2. Funding research and development

The government can allocate funding to support research and development of AI systems for disaster management that prioritize ethical considerations. Such funding could be directed towards research in areas such as bias mitigation, transparency, and accountability, as well as towards the development of open-source software tools that facilitate ethical AI development.

However, this approach is not without challenges. One such challenge is limited funding, as the government may not have adequate resources to finance all necessary ethical AI research and development in disaster management, making it challenging to prioritize the areas of focus. Another challenge is uncertainty about the outcomes, due to the unpredictable nature of AI research and development, with no guarantee that ethical AI systems will be successfully developed or implemented. Lack of collaboration between government, AI developers, disaster management experts, and other stakeholders is another challenge, potentially leading to the creation of AI systems that do not meet the needs of disaster management and that could result in unintended consequences. Furthermore, the rapid evolution of AI technology causes a risk of overlooking ethical considerations in the race to develop new technology. Regulatory challenges are also common, and the government may struggle to create regulations that are effective in promoting ethical AI development while also not stifling innovation. Finally, ethical considerations may not be prioritized. In some cases, the focus may be more on developing AI systems that are effective in managing disasters, instead of ensuring that they are ethically sound.

#### 4.1.3. Building Capacity

The government can support capacity-building efforts to ensure that those working in disaster management have the requisite knowledge and skills for handling AI systems in an ethical and responsible manner. This could include funding for training projects, workshops, and other educational initiatives.

This solution also faces challenges related to research funding. Additionally, government agencies may resist change, hindering the implementation of capacity-building initiatives due to entrenched bureaucratic structures and a fear of disruption[133]

#### 4.1.4. Monitoring and Evaluation

The government can implement monitoring and evaluation processes to ensure that AI systems used in disaster management operate in an ethical and responsible manner. This can involve the development of frameworks for monitoring and evaluating AI systems, as well as the establishment of oversight mechanisms to ensure that AI systems are operating in conformity to ethical guidelines and standards.

### 4.2. Embedded ethics: possible drawbacks

While embedding ethics directly into AI systems is a potential solution to prevent and alleviate the ethical repercussions of AI use in disaster management, it also faces several challenges and deficits.

The first problem is the challenge of defining ethics. Determining ethical principles and embedding them into AI systems can be complex and difficult. Stakeholders may have varying opinions on what constitutes ethical behavior and how to translate ethical principles into concrete rules for AI systems. This is linked to the problem of handling ethical trade-offs as well. Furthermore, current ethical frameworks may not fully address the unique ethical challenges presented by AI systems in disaster management.

However, the problem might be rendered less severe if common ethical principles are taken into account, which engage multiple stakeholders (including experts in ethics, AI developers, policymakers, and members of affected communities) and are agreed upon by people from diverse cultures and backgrounds. The process should be iterative, involving continually revisiting and refining ethical principles based on new developments in AI technology and feedback from stakeholders. Furthermore, an assessment of ethical impacts should be carried out to identify potential ethical risks and challenges associated with AI systems, which will inform the development of ethical principles. Transparency and accountability must be built into AI systems to ensure that ethical principles are upheld and that AI systems are behaving in a responsible and ethical manner. This can be achieved through measures such as explainability, auditing, and human oversight.

The second problem is difficulty in implementation. Embedding ethical principles into AI systems can be a difficult and time-consuming process, particularly for complex systems with numerous variables and inputs. It may also require specialized expertise in both AI development and ethics. For instance, it is argued that practical challenges—such as inconsistent regulatory pressures, conflicting business goals, data quality issues, and system integration complexities—can hinder the effective implementation of ethical AI[134].

To overcome this difficulty, the following measures can be helpful: (1) clear guidance and standards for AI developers on how to implement ethical principles, which can include creating industry-wide standards or guidelines for ethical AI development, or providing specific instructions for developers working on projects with ethical implications. (2) Training and education for developers on how to implement ethical principles into AI systems, which can help ensure that they have the requisite knowledge and skills to do so. This can include training on ethical principles and guidelines, as well as technical training on how to implement these principles into AI systems. (3) Collaboration with stakeholders, such as affected communities, regulators, and ethical experts. This also can help ensure that ethical principles are effectively implemented in AI systems. In this way, it is guaranteed that diverse perspectives are taken into account, and potential implementation challenges and solutions can be identified. (4) Testing and evaluation AI systems for their adherence to ethical principles. This can help identify areas for improvement and ensure that these principles are effectively implemented. This includes both technical testing, such as testing for bias or transparency, as well as testing by human evaluators. (5) Incentives and rewards for developers who effectively implement ethical principles in AI systems. This can help promote ethical behavior and ensure that ethical principles are prioritized in AI development.

Finally, there is a risk of unintended consequences. Embedding ethical principles into AI systems may have unintended consequences, particularly if the ethical principles are not well-defined or are not effectively implemented. There is a risk that AI systems may not behave as intended, or may even have unintended negative consequences. Examples include the potential negative impacts of machine learning technologies in disaster risk management, emphasizing the need for careful assessment of ethical, fairness, and bias considerations[135].

The risk can be mitigated thorough testing and evaluation, which can help identify potential unintended consequences and ensure that these systems are behaving as intended. This can involve both technical testing, such as testing for accuracy and reliability, and testing by human evaluators. Another mitigating factor is anticipatory governance, which involves anticipating potential unintended consequences of AI systems and designing governance frameworks to address these risks. This includes engaging with stakeholders to identify potential risks, developing policies and regulations to deal with these risks, and establishing mechanisms for continuous monitoring and evaluation.

It is also helpful to develop transparency and accountability as well as responsible practices. For instance, building diversity and inclusion into the design process can help mitigate the risk of unintended consequences by ensuring that a wide range of perspectives and experiences are taken into account.

### 4.3. Accountability and Liability of AI-caused Disasters

Determining accountability for AI-caused disasters or AI-caused disaster mismanagement can be complex, as it can involve multiple actors and factors. In general, human individuals or organizations tend to be held accountable for AI-caused disasters. For instance, AI developers might be deemed culprits when they create AI systems that are used in disaster management and the system causes harm due to design flaws, insufficient testing, or failure to take account of ethical implications. Sometimes AI operators are blamed. Operators who are responsible for implementing and maintaining AI systems in disaster management can be held accountable if they fail to properly operate and maintain the system, or if they use the system in a way that causes harm. Regulators are also blamed under certain circumstances. Regulators who oversee the development and use of AI systems for disaster management can be held accountable if they fail to develop and enforce appropriate regulations and standards, or if they fail to adequately monitor and address issues with AI systems. Further, governments might be deemed culprit. They may be held accountable if they fail to provide appropriate funding, oversight, or support for the development and use of ethical AI systems in disaster management. And finally, other stakeholders, such as affected communities, NGOs, and disaster response agencies, can be held accountable if they contribute to or fail to address issues with the use of AI in disaster management [see 115].

It might be argued, however, that AI systems themselves might also be held accountable under certain circumstances if other human agents have done their tasks properly, but the AI has gone through self-learning phases and made decisions which caused the disaster, but cannot be attributed to its developers. In this case, we can say that AI has some sort of "semi-agency." The challenge, nevertheless, is how to hold an AI system accountable. Individuals or organizations can be held accountable for their misconducts or for the disasters they cause by being fined or by some of their rights being suspended. At present, AI systems do not enjoy any rights or legal protections that might be suspended. However, if AI systems become more and more autonomous, then they might be granted certain rights, which might then be suspended as a way of punishing them for their misconducts. The question of the sentience or consciousness of large language models or other sophisticated AI systems might also have an impact on whether they have any rights and whether they might be held accountable [104].

## 5. Conclusion and Future Directions

There are several areas of research into an ethical IDM that require further exploration. For instance, although fairness in AI systems for disaster management is an emerging concern, further dedicated research is needed to develop specific techniques tailored to these contexts. Therefore, future research could focus on developing techniques to detect and mitigate bias in disaster management AI systems to secure fair and equitable outcomes.

In the context of disaster management, it may be of significance to explore how resilience and adaptability can enhance the performance of AI systems in evolving disaster contexts, as the field is constantly changing.

Future research may explore the ways in which AI systems can be designed and developed to exhibit resilience and adaptability, allowing them to learn from new experiences and data.

Human-robot interaction is another area of research that is in need of further investigation. While robots and autonomous AI systems are increasingly deployed in disaster management, more targeted studies are needed on how humans and robots can effectively collaborate in such scenarios. As such, future research could focus on developing strategies to facilitate effective communication and collaboration between humans and robots.

In disaster management, decision-making tends to be collaborative, involving emergency responders, policymakers, and other stakeholders. Future research may explore how AI systems can support collaborative decision-making in these situations. For example, AI systems could provide real-time information and predictions to facilitate the decision-making process.
Finally, long-term impacts should be taken into account when addressing disaster management. Future research could examine how AI systems can aid in long-term recovery efforts by analyzing data on the social, economic, and environmental impacts of disasters. Moreover, AI systems could provide recommendations for long-term planning and recovery efforts.

**Conflict of interest**: The authors declare that the research was conducted in the absence of any commercial or financial relationships that could be construed as a potential conflict of interest.

**Acknowledgements:** In accordance with MLA (Modern Language Association) guidelines, we note the use AI-powered tools, such as OpenAI's applications, for assistance in editing and brainstorming.

## References

[1] H. ; R. M. Ritchie, "Natural Disasters," 2014.

[2] World Meteorological Organization, "WMO ATLAS OF MORTALITY AND ECONOMIC LOSSES FROM WEATHER, CLIMATE AND WATER EXTREMES (1970–2019)," 2021.

[3] deepdesignlab, "Disaster," Jul. 2021.

[4] A. Norris, "Disaster E-Health: A New Paradigm for Collaborative Healthcare in Disasters," *ISCRAM*, 2015.

[5] D. Guha-Sapir, "Thirty years of natural disasters 1974-2003: The numbers," *Presses univ. de Louvain*, 2004.

[6] I. M. Shaluf, "An overview on disasters," *Disaster Prevention and Management: An International Journal* , 2007.

[7] WorldOMeters, "https://www.worldometers.info/coronavirus/," May 2022.

[8] World Health Organization (WHO), "14.9 million excess deaths associated with the COVID-19 pandemic in 2020 and 2021," May 2022.

[9] A. et al. Liberati, "The PRISMA statement for reporting systematic reviews and meta-analyses of studies that evaluate health care interventions: explanation and elaboration," *Journal of clinical epidemiology* , 2009.

[10] W. Sun, P. Bocchini, and B. D. Davison, "Applications of artificial intelligence for disaster management," *Natural Hazards*, vol. 103, no. 3, pp. 2631–2689, 2020, doi: 10.1007/s11069-020-04124-3.

[11] Z. Ashktorab, "Tweedr: Mining twitter to inform disaster response. ," *ISCRAM*, 2014.

[12] S. K. Abid *et al.*, "Toward an Integrated Disaster Management Approach: How Artificial Intelligence Can Boost Disaster Management," *Sustainability*, vol. 13, no. 22, p. 12560, 2021, doi: 10.3390/su132212560.

[13] M. Schofield, "An Artificial Intelligence (AI) Approach to Controlling Disaster Scenarios," *Future Role of Sustainable Innovative Technologies in Crisis Management*, 2022.

[14] M. Srivastava, T. Abdelzaher, and B. Szymanski, "Human-centric sensing," *Philosophical Transactions of the Royal Society A: Mathematical, Physical and Engineering Sciences*, vol. 370, no. 1958, pp. 176–197, 2012, doi: 10.1098/rsta.2011.0244.

[15] M. Goodchild and Alan Glennon, "Crowdsourcing geographic information for disaster response: a research frontier," *Int J Digit Earth*, 2010.

[16] M. S. Kiatpanont Rungsun, P. Tanlamai Uthai, and P. Chongstitvatana Prabhas, "Extraction of actionable information from crowdsourced disaster data," *Journal of Emergency Management*, vol. 14, no. 6, p. 377, 2016, doi: 10.5055/jem.2016.0302.

[17] S. Kim, H. Kim, and Y. Namkoong, "Ordinal Classification of Imbalanced Data with Application in Emergency and Disaster Information Services," *IEEE Intell Syst*, vol. 31, no. 5, pp. 50–56, 2016, doi: 10.1109/MIS.2016.27.

[18] H. R. Rasouli, H. R. Zahedi, M. Abbasi Farajzadeh, A. Aliakbar Esfahani, and F. Ahmadpour, "Medical Aspects of Earthquakes in Iran," *Trauma Mon*, vol. 23, no. 5, 2018, doi: 10.5812/traumamon.80528.

[19] F. Cavdur, A. Sebatli-Saglam, and M. Kose-Kucuk, "A spreadsheet-based decision support tool for temporary-disaster-response facilities allocation," *Saf Sci*, vol. 124, p. 104581, 2020, doi: 10.1016/j.ssci.2019.104581.

[20] R. Ba, Q. Deng, Y. Liu, R. Yang, and H. Zhang, "Multi-hazard disaster scenario method and emergency management for urban resilience by integrating experiment–simulation–field data," *Journal of Safety Science and Resilience*, vol. 2, no. 2, pp. 77–89, 2021, doi: 10.1016/j.jnlssr.2021.05.002.

[21] S. K. Abid *et al.*, "Toward an Integrated Disaster Management Approach: How Artificial Intelligence Can Boost Disaster Management," *Sustainability*, vol. 13, no. 22, p. 12560, 2021, doi: 10.3390/su132212560.

[22] L. Tan, Ji Guo, Selvarajah Mohanarajah, and Kun Zhou, "Can we detect trends in natural disaster management with artificial intelligence? A review of modeling practices," *Natural Hazards* , 2021.

[23] A. M. J. Skulimowski and V. A. Bañuls, "AI Alignment of Disaster Resilience Management Support Systems," L. Rutkowski, R. Scherer, M. Korytkowski, W. Pedrycz, R. Tadeusiewicz, and

J. M. Zurada, Eds., in Lecture Notes in Computer Science. Springer International Publishing, 2021, pp. 354–366. doi: 10.1007/978-3-030-87897-9_32.

[24] T. Yabe, P. S. C. Rao, S. v Ukkusuri, and S. L. Cutter, "Toward data-driven, dynamical complex systems approaches to disaster resilience," *Proceedings of the National Academy of Sciences*, vol. 119, no. 8, p. e2111997119, 2022, doi: 10.1073/pnas.2111997119.

[25] M. Mendoza, B. Poblete, and C. Castillo, "Twitter under crisis: can we trust what we RT?," in SOMA '10. Association for Computing Machinery, 2010, pp. 71–79. doi: 10.1145/1964858.1964869.

[26] J. Li and H. r. Rao, "Twitter as a Rapid Response News Service: An Exploration in the Context of the 2008 China Earthquake," *THE ELECTRONIC JOURNAL OF INFORMATION SYSTEMS IN DEVELOPING COUNTRIES*, vol. 42, no. 1, pp. 1–22, 2010, doi: 10.1002/j.1681-4835.2010.tb00300.x.

[27] B. Takahashi, E. C. Tandoc, and C. Carmichael, "Communicating on Twitter during a disaster: An analysis of tweets during Typhoon Haiyan in the Philippines," *Comput Human Behav*, vol. 50, pp. 392–398, 2015, doi: 10.1016/j.chb.2015.04.020.

[28] F. Laylavi, A. Rajabifard, and M. Kalantari, "Event relatedness assessment of Twitter messages for emergency response," *Inf Process Manag*, vol. 53, no. 1, pp. 266–280, 2017, doi: 10.1016/j.ipm.2016.09.002.

[29] D. Reynard and M. Shirgaokar, "Harnessing the power of machine learning: Can Twitter data be useful in guiding resource allocation decisions during a natural disaster?," *Transp Res D Transp Environ*, vol. 77, pp. 449–463, 2019, doi: 10.1016/j.trd.2019.03.002.

[30] Y. Rizk, H. S. Jomaa, M. Awad, and C. Castillo, "A computationally efficient multi-modal classification approach of disaster-related Twitter images," in SAC '19. Association for Computing Machinery, 2019, pp. 2050–2059. doi: 10.1145/3297280.3297481.

[31] P. Kumar, F. Ofli, M. Imran, and C. Castillo, "Detection of Disaster-Affected Cultural Heritage Sites from Social Media Images Using Deep Learning Techniques," *Journal on Computing and Cultural Heritage*, vol. 13, no. 3, pp. 23:1-23:31, 2020, doi: 10.1145/3383314.

[32] V. v Mihunov, N. S. N. Lam, L. Zou, Z. Wang, and K. Wang, "Use of Twitter in disaster rescue: lessons learned from Hurricane Harvey," *Int J Digit Earth*, vol. 13, no. 12, pp. 1454–1466, 2020, doi: 10.1080/17538947.2020.1729879.

[33] R. Perry, "Defining Disaster: An Evolving Concept. In Handbook of disaster research," *Springer*, 2018.

[34] K. Mase, "How to deliver your message from/to a disaster area," *IEEE Communications Magazine*, vol. 49, no. 1, pp. 52–57, 2011, doi: 10.1109/MCOM.2011.5681015.

[35] C. M. Cherian, N. Jayaraj, and S. G. Vaidyanathan, "Artificially Intelligent Tsunami Early Warning System," in *2010 12th International Conference on Computer Modelling and Simulation*, 2010, pp. 39–44. doi: 10.1109/UKSIM.2010.16.

[36] R. Lamsal and T. V. Kumar, "Artificial Intelligence and Early Warning Systems," *AI and Robotics in Disaster Studies. Palgrave Macmillan, Singapore*, 2020.

[37] C. l. Wu and K. w. Chau, “A flood forecasting neural network model with genetic algorithm,” *Int J Environ Pollut*, vol. 28, no. 3–4, pp. 261–273, 2006, doi: 10.1504/IJEP.2006.011211.

[38] M.-H. Huang and R. T. Rust, “Artificial Intelligence in Service,” *J Serv Res*, vol. 21, no. 2, pp. 155–172, 2018, doi: 10.1177/1094670517752459.

[39] P. et al. Mittal, “Dual artificial neural network for rainfall-runoff forecasting,” *J Water Resour Prot*, 2012.

[40] I. K. , et al. Hadihardaja, “Decision support system for predicting flood characteristics based on database modelling development (case study: Upper Citarum, West Java, Indonesia),” *WIT Transactions on Ecology and the Environment* , 2012.

[41] F. H. et al. Chowdhury, “Design, control & performance analysis of forecast junction IoT and swarm robotics based system for natural disaster monitoring,” *8th International Conference on Computing, Communication and Networking Technologies (ICCCNT). IEEE*, 2017.

[42] E. et al. Danso-Amoako, “Predicting dam failure risk for sustainable flood retention basins: A generic case study for the wider Greater Manchester area,” *Comput Environ Urban Syst*, 2012.

[43] K. Nagatani *et al.*, “Emergency response to the nuclear accident at the Fukushima Daiichi Nuclear Power Plants using mobile rescue robots,” *J Field Robot*, vol. 30, no. 1, pp. 44–63, 2013, doi: 10.1002/rob.21439.

[44] F. A. Ruslan, Zainazlan Md Zain, and Ramli Adnan, “Flood prediction using NARX neural network and EKF prediction technique: A comparative study,” *IEEE 3rd International Conference on System Engineering and Technology*, 2013.

[45] A. P. Anindita, Pujo Laksono, Gusti Bagus, and Baskara Nugraha, “Dam water level prediction system utilizing Artificial Neural Network Back Propagation: Case study: Ciliwung watershed, Katulampa Dam,” *International Conference on ICT For Smart Society (ICISS). IEEE*, 2016.

[46] R. Adnan, F. A. Ruslan, A. M. Samad, and Z. M. Zain, “Artificial neural network modelling and flood water level prediction using extended Kalman filter,” in *2012 IEEE International Conference on Control System, Computing and Engineering*, 2012, pp. 535–538. doi: 10.1109/ICCSCE.2012.6487204.

[47] K. Kochersberger, K. Kroeger, B. Krawiec, E. Brewer, and T. Weber, “Post-disaster Remote Sensing and Sampling via an Autonomous Helicopter,” *J Field Robot*, vol. 31, no. 4, pp. 510–521, 2014, doi: 10.1002/rob.21502.

[48] J. E. DeVault, “Robotic system for underwater inspection of bridge piers,” *IEEE Instrumentation Measurement Magazine*, vol. 3, no. 3, pp. 32–37, 2000, doi: 10.1109/5289.863909.

[49] C. T. Recchiuto and Antonio Sgorbissa, “Post-disaster assessment with unmanned aerial vehicles: A survey on practical implementations and research approaches,” *Journal of Field Robotics* , 2018.

[50] M. M. Torok, M. Golparvar-Fard, and K. B. Kochersberger, “Image-Based Automated 3D Crack Detection for Post-disaster Building Assessment,” *Journal of Computing in Civil Engineering*, vol. 28, no. 5, p. A4014004, 2014, doi: 10.1061/(ASCE)CP.1943-5487.0000334.

[51] R. R. Murphy, J. Kravitz, S. L. Stover, and R. Shoureshi, “Mobile robots in mine rescue and recovery,” *IEEE Robotics Automation Magazine*, vol. 16, no. 2, pp. 91–103, 2009, doi: 10.1109/MRA.2009.932521.

[52] E. T. Steimle, R. R. Murphy, M. Lindemuth, and M. L. Hall, “Unmanned marine vehicle use at Hurricanes Wilma and Ike,” in *OCEANS 2009*, 2009, pp. 1–6. doi: 10.23919/OCEANS.2009.5422201.

[53] S. Fang *et al.*, “An integrated information system for snowmelt flood early-warning based on internet of things,” *Information Systems Frontiers*, vol. 17, no. 2, pp. 321–335, 2015, doi: 10.1007/s10796-013-9466-1.

[54] J. et al Cen, “Developing a disaster surveillance system based on wireless sensor network and cloud platform,” *IET International Conference on Communication Technology and Application (ICCTA 2011)*, 2011.

[55] Q. et al. Huang, “DisasterMapper: A CyberGIS framework for disaster management using social media data,” *Proceedings of the 4th International ACM SIGSPATIAL Workshop on Analytics for Big Geospatial Data*, 2015.

[56] Z. Allam, Gourav Dey, and David S. Jones, “Artificial intelligence (AI) provided early detection of the coronavirus (COVID-19) in China and will influence future Urban health policy internationally,” *Ai*, vol. 1, no. 2, 2020.

[57] S. Afroogh *et al.*, “Tracing app technology: an ethical review in the COVID-19 era and directions for post-COVID-19,” *Ethics Inf Technol*, vol. 24, no. 3, p. 30, 2022, doi: 10.1007/s10676-022-09659-6.

[58] A. Dangwal, “Ukraine Uses ‘Controversial’ Artificial Intelligence Tech In Its War Against Russia As Kiev Looks To Win The ‘Digital War,’” *The Guardian*, Apr. 2022.

[59] D. A. , Gill and L. A. Ritchie, “Contributions of technological and natech disaster research to the social science disaster paradigm. In Handbook of disaster research (pp. 39-60),” *Springer*, 2018.

[60] M. Gähler, “Remote Sensing for Natural or Man-made Disasters and Environmental Changes,” *Environmental applications of remote sensing. InTech*, 2016.

[61] R. Thomson *et al.*, “Trusting Tweets: The Fukushima Disaster and Information Source Credibility on Twitter,” p. 10, 2012, [Online]. Available: files/8262/Thomson et al. - 2012 - Trusting Tweets The Fukushima Disaster and Inform.pdf

[62] P. Dialani, “DIFFERENT WAYS OF HOW TWITTER USES ARTIFICIAL INTELLIGENCE,” *ARTIFICIAL INTELLIGENCE LATEST NEWS*, Jan. 2019.

[63] B. et al. Baruque, “A forecasting solution to the oil spill problem based on a hybrid intelligent system,” *Information Sciences* , 2010.

[64] S.-K. Kim and Jun-Ho Huh, “Blockchain of carbon trading for UN sustainable development goals,” *Sustainability*, 2020.

[65] O. I. Olayode, L. K. Tartibu, and M. O. Okwu, “Application of Artificial Intelligence in Traffic Control System of Non-autonomous Vehicles at Signalized Road Intersection.,” *Procedia CIRP*, vol. 91, pp. 194–200, 2020, doi: 10.1016/j.procir.2020.02.167.

[66] A. I. Canhoto, “Leveraging machine learning in the global fight against money laundering and terrorism financing: An affordances perspective,” *J Bus Res*, 2021.

[67] J. Johnson, “Artificial intelligence & future warfare: implications for international security.,” *Defense & Security Analysis 35.2*, 2019.

[68] B. et al. Hallaq, “Artificial intelligence within the military domain and cyber warfare,” *Eur. Conf. Inf. Warf. Secur. ECCWS*, 2017.

[69] F. E. Morgan *et al.*, “Military Applications of Artificial Intelligence: Ethical Concerns in an Uncertain World,” RAND PROJECT AIR FORCE SANTA MONICA CA SANTA MONICA United States, 2020. [Online]. Available: https://apps.dtic.mil/sti/citations/AD1097313

[70] P. M. Duggan, “Strategic Development of Special Warfare in Cyberspace,” p. 8, 2015, [Online]. Available: files/8290/Duggan - 2015 - Strategic Development of Special Warfare in Cybers.pdf

[71] M. Maybury, “Toward the Assured Cyberspace Advantage: Air Force Cyber Vision 2025,” *IEEE Security Privacy*, vol. 13, no. 1, pp. 49–56, 2015, doi: 10.1109/MSP.2013.135.

[72] D. D. Djordjevich, P. G. Xavier, M. L. Bernard, J. H. Whetzel, M. R. Glickman, and S. J. Verzi, “Preparing for the aftermath: Using emotional agents in game-based training for disaster response,” in *2008 IEEE Symposium On Computational Intelligence and Games*, 2008, pp. 266–275. doi: 10.1109/CIG.2008.5035649.

[73] S. Ali, P. Biermanns, R. Haider, and K. Reicherter, “Landslide susceptibility mapping by using a geographic information system (GIS) along the China–Pakistan Economic Corridor (Karakoram Highway), Pakistan,” *Natural Hazards and Earth System Sciences*, vol. 19, no. 5, pp. 999–1022, 2019, doi: 10.5194/nhess-19-999-2019.

[74] B. Takahashi, E. C. Tandoc, and C. Carmichael, “Communicating on Twitter during a disaster: An analysis of tweets during Typhoon Haiyan in the Philippines,” *Comput Human Behav*, vol. 50, pp. 392–398, 2015, doi: 10.1016/j.chb.2015.04.020.

[75] M. Mendoza, B. Poblete, and C. Castillo, “Twitter under crisis: can we trust what we RT?,” in SOMA ’10. Association for Computing Machinery, 2010, pp. 71–79. doi: 10.1145/1964858.1964869.

[76] J. Li and H. r. Rao, “Twitter as a Rapid Response News Service: An Exploration in the Context of the 2008 China Earthquake,” *THE ELECTRONIC JOURNAL OF INFORMATION SYSTEMS IN DEVELOPING COUNTRIES*, vol. 42, no. 1, pp. 1–22, 2010, doi: 10.1002/j.1681-4835.2010.tb00300.x.

[77] D. Reynard and M. Shirgaokar, “Harnessing the power of machine learning: Can Twitter data be useful in guiding resource allocation decisions during a natural disaster?,” *Transp Res D Transp Environ*, vol. 77, pp. 449–463, 2019, doi: 10.1016/j.trd.2019.03.002.

[78] Y. Rizk, H. S. Jomaa, M. Awad, and C. Castillo, “A computationally efficient multi-modal classification approach of disaster-related Twitter images,” in SAC ’19. Association for Computing Machinery, 2019, pp. 2050–2059. doi: 10.1145/3297280.3297481.

[79] V. v Mihunov, N. S. N. Lam, L. Zou, Z. Wang, and K. Wang, “Use of Twitter in disaster rescue: lessons learned from Hurricane Harvey,” *Int J Digit Earth*, vol. 13, no. 12, pp. 1454–1466, 2020, doi: 10.1080/17538947.2020.1729879.

[80] G. A. Ruz, P. A. Henríquez, and A. Mascareño, “Sentiment analysis of Twitter data during critical events through Bayesian networks classifiers,” *Future Generation Computer Systems*, vol. 106, pp. 92–104, 2020, doi: 10.1016/j.future.2020.01.005.

[81] S. Saravi, R. Kalawsky, D. Joannou, M. Rivas Casado, G. Fu, and F. Meng, “Use of Artificial Intelligence to Improve Resilience and Preparedness Against Adverse Flood Events,” *Water (Basel)*, vol. 11, no. 5, p. 973, 2019, doi: 10.3390/w11050973.

[82] F. , D. G. and B. Hétu. Gauthier, “Logistic models as a forecasting tool for snow avalanches in a cold maritime climate: northern Gaspésie, Québec, Canada,” *Natural hazards*, vol. 89, no. 1, 2017.

[83] A. C. Teodoro and L. Duarte, “Forest fire risk maps: a GIS open source application – a case study in Norwest of Portugal,” *International Journal of Geographical Information Science*, vol. 27, no. 4, pp. 699–720, 2013, doi: 10.1080/13658816.2012.721554.

[84] K. Mase, “How to deliver your message from/to a disaster area,” *IEEE Communications Magazine*, vol. 49, no. 1, pp. 52–57, 2011, doi: 10.1109/MCOM.2011.5681015.

[85] M. Rauter and D. Winkler, “Predicting Natural Hazards with Neuronal Networks,” arXiv, 2018. [Online]. Available: http://arxiv.org/abs/1802.07257

[86] B. Choubin, M. Borji, A. Mosavi, F. Sajedi-Hosseini, V. P. Singh, and S. Shamshirband, “Snow avalanche hazard prediction using machine learning methods,” *J Hydrol (Amst)*, vol. 577, p. 123929, 2019, doi: 10.1016/j.jhydrol.2019.123929.

[87] J. Huang, X. Wang, Y. Zhao, C. Xin, and H. Xiang, “LARGE EARTHQUAKE MAGNITUDE PREDICTION IN TAIWAN BASED ON DEEP LEARNING NEURAL NETWORK,” *Neural Network World*, vol. 28, no. 2, pp. 149–160, 2018, doi: 10.14311/NNW.2018.28.009.

[88] D. T. , Bui, N. D. Hoang, F. Martínez-Álvarez, P. T. T. Ngo, P. v. Hoa, and T. D. Pham, “A novel deep learning neural network approach for predicting flash flood susceptibility: A case study at a high frequency tropical storm area,” *Science of The Total Environment*, 2020.

[89] R. Adnan, F. A. Ruslan, A. M. Samad, and Z. M. Zain, “Artificial neural network modelling and flood water level prediction using extended Kalman filter,” in *2012 IEEE International Conference on Control System, Computing and Engineering*, 2012, pp. 535–538. doi: 10.1109/ICCSCE.2012.6487204.

[90] S. Ali, P. Biermanns, R. Haider, and K. Reicherter, “Landslide susceptibility mapping by using a geographic information system (GIS) along the China–Pakistan Economic Corridor (Karakoram Highway), Pakistan,” *Natural Hazards and Earth System Sciences*, vol. 19, no. 5, pp. 999–1022, 2019, doi: 10.5194/nhess-19-999-2019.

[91] F. Cavdur, A. Sebatli-Saglam, and M. Kose-Kucuk, “A spreadsheet-based decision support tool for temporary-disaster-response facilities allocation,” *Saf Sci*, vol. 124, p. 104581, 2020, doi: 10.1016/j.ssci.2019.104581.

[92] S. Fang *et al.*, "An integrated information system for snowmelt flood early-warning based on internet of things," *Information Systems Frontiers*, vol. 17, no. 2, pp. 321–335, 2015, doi: 10.1007/s10796-013-9466-1.

[93] C. Roy, "An Informed System Development Approach to Tropical Cyclone Track and Intensity Forecasting," 2016, [Online]. Available: http://urn.kb.se/resolve?urn=urn:nbn:se:liu:diva-123198

[94] C. M. Cherian, N. Jayaraj, and S. G. Vaidyanathan, "Artificially Intelligent Tsunami Early Warning System," in *2010 12th International Conference on Computer Modelling and Simulation*, 2010, pp. 39–44. doi: 10.1109/UKSIM.2010.16.

[95] R. Chen, Weimin Zhang, and Xiang Wang, "Machine learning in tropical cyclone forecast modeling: A review," *Atmosphere (Basel)*, vol. 11, no. 7, 2020.

[96] A. Mosavi, P. Ozturk, and K. Chau, "Flood Prediction Using Machine Learning Models: Literature Review," *Water (Basel)*, vol. 10, no. 11, p. 1536, 2018, doi: 10.3390/w10111536.

[97] S. Jeong and D. K. Yoon., "Examining vulnerability factors to natural disasters with a spatial autoregressive model: The case of South Korea," *Sustainability*, vol. 10, no. 5, 2018.

[98] N. Yoon and H.-K. Lee, "AI Recommendation Service Acceptance: Assessing the Effects of Perceived Empathy and Need for Cognition," *Journal of Theoretical and Applied Electronic Commerce Research*, vol. 16, no. 5, pp. 1912–1928, 2021, doi: 10.3390/jtaer16050107.

[99] C. l. Wu and K. w. Chau, "A flood forecasting neural network model with genetic algorithm," *Int J Environ Pollut*, vol. 28, no. 3–4, pp. 261–273, 2006, doi: 10.1504/IJEP.2006.011211.

[100] M. A. Nabian and H. Meidani, "Deep Learning for Accelerated Seismic Reliability Analysis of Transportation Networks," *Computer-Aided Civil and Infrastructure Engineering*, vol. 33, no. 6, pp. 443–458, 2018, doi: 10.1111/mice.12359.

[101] W. Sun, P. Bocchini, and B. D. Davison, "Applications of artificial intelligence for disaster management," *Natural Hazards*, vol. 103, no. 3, pp. 2631–2689, 2020, doi: 10.1007/s11069-020-04124-3.

[102] A. Neshat, B. Pradhan, and M. Dadras, "Groundwater vulnerability assessment using an improved DRASTIC method in GIS," *Resour Conserv Recycl*, vol. 86, pp. 74–86, 2014, doi: 10.1016/j.resconrec.2014.02.008.

[103] B. Ortiz *et al.*, "Improving Community Resiliency and Emergency Response With Artificial Intelligence," *arXiv:2005.14212 [cs]*, 2020, [Online]. Available: http://arxiv.org/abs/2005.14212

[104] R. Ba, Q. Deng, Y. Liu, R. Yang, and H. Zhang, "Multi-hazard disaster scenario method and emergency management for urban resilience by integrating experiment–simulation–field data," *Journal of Safety Science and Resilience*, vol. 2, no. 2, pp. 77–89, 2021, doi: 10.1016/j.jnlssr.2021.05.002.

[105] I. Mitsopoulos and G. Mallinis, "A data-driven approach to assess large fire size generation in Greece," *Natural Hazards*, vol. 88, no. 3, pp. 1591–1607, 2017, doi: 10.1007/s11069-017-2934-z.

[106] J. Cen, T. Yu, Z. Li, S. Jin, and S. Liu, "Developing a disaster surveillance system based on wireless sensor network and cloud platform," in *IET International Conference on Communication Technology and Application (ICCTA 2011)*, 2011, pp. 757–761. doi: 10.1049/cp.2011.0770.

[107] J. E. DeVault, “Robotic system for underwater inspection of bridge piers,” *IEEE Instrumentation Measurement Magazine*, vol. 3, no. 3, pp. 32–37, 2000, doi: 10.1109/5289.863909.

[108] R. R. Murphy, J. Kravitz, S. L. Stover, and R. Shoureshi, “Mobile robots in mine rescue and recovery,” *IEEE Robotics Automation Magazine*, vol. 16, no. 2, pp. 91–103, 2009, doi: 10.1109/MRA.2009.932521.

[109] M. Imran, C. Castillo, J. Lucas, P. Meier, and S. Vieweg, “AIDR: artificial intelligence for disaster response,” in WWW ’14 Companion. Association for Computing Machinery, 2014, pp. 159–162. doi: 10.1145/2567948.2577034.

[110] K. et al. Nagatani, “Volcanic ash observation in active volcano areas using teleoperated mobile robots-Introduction to our robotic-volcano-observation project and field experiments,” *IEEE International Symposium on Safety, Security, and Rescue Robotics (SSRR)*, 2013.

[111] M. E. Mohammadi, Daniel P. Watson, and Richard L. Wood., “Deep learning-based damage detection from aerial SfM point clouds,” *Drones*, vol. 3, no. 3, 2009.

[112] F. Laylavi, A. Rajabifard, and M. Kalantari, “Event relatedness assessment of Twitter messages for emergency response,” *Inf Process Manag*, vol. 53, no. 1, pp. 266–280, 2017, doi: 10.1016/j.ipm.2016.09.002.

[113] S. Kuang and B. D. Davison, “Learning Word Embeddings with Chi-Square Weights for Healthcare Tweet Classification,” *Applied Sciences*, vol. 7, no. 8, p. 846, 2017, doi: 10.3390/app7080846.

[114] M. E. Aydin and R. Fellows, “Building Collaboration in Multi-agent Systems Using Reinforcement Learning,” N. T. Nguyen, E. Pimenidis, Z. Khan, and B. Trawiński, Eds., in Lecture Notes in Computer Science. Springer International Publishing, 2018, pp. 201–212. doi: 10.1007/978-3-319-98446-9_19.

[115] K. Kochersberger, K. Kroeger, B. Krawiec, E. Brewer, and T. Weber, “Post-disaster Remote Sensing and Sampling via an Autonomous Helicopter,” *J Field Robot*, vol. 31, no. 4, pp. 510–521, 2014, doi: 10.1002/rob.21502.

[116] M. M. Torok, M. Golparvar-Fard, and K. B. Kochersberger, “Image-Based Automated 3D Crack Detection for Post-disaster Building Assessment,” *Journal of Computing in Civil Engineering*, vol. 28, no. 5, p. A4014004, 2014, doi: 10.1061/(ASCE)CP.1943-5487.0000334.

[117] L. Zou, N. S. N. Lam, H. Cai, and Y. Qiang, “Mining Twitter Data for Improved Understanding of Disaster Resilience,” *Ann Am Assoc Geogr*, vol. 108, no. 5, pp. 1422–1441, 2018, doi: 10.1080/24694452.2017.1421897.

[118] W. T. Chen and Ying-Hua Huang, “Approximately predicting the cost and duration of school reconstruction projects in Taiwan,” *Construction management and Economics*, vol. 24, no. 12, 2006.

[119] A. T. Zagorecki, D. E. A. Johnson, and J. Ristvej, “Data mining and machine learning in the context of disaster and crisis management,” *International Journal of Emergency Management*, vol. 9, no. 4, pp. 351–365, 2013, doi: 10.1504/IJEM.2013.059879.

[120] T. Yabe, P. S. C. Rao, S. v Ukkusuri, and S. L. Cutter, "Toward data-driven, dynamical complex systems approaches to disaster resilience," *Proceedings of the National Academy of Sciences*, vol. 119, no. 8, p. e2111997119, 2022, doi: 10.1073/pnas.2111997119.

[121] M. Keim and Paul Giannone, "Disaster preparedness," *Disaster Medicine, Philadelphia, PA: Mosby* , pp. 164–173, 2006.

[122] Washington Post, "Teslas running Autopilot involved in 273 crashes reported since last year," Jul. 2022.

[123] James Vincent, "Twitter taught Microsoft's AI chatbot to be a racist asshole in less than a day," *www.theverge.com*, Mar. 2016.

[124] The New York Times, "Driver Charged in Uber's Fatal 2018 Autonomous Car Crash," Dec. 2020.

[125] S. Afroogh, A. Kazemi, and A. Seyedkazemi, "COVID-19, Scarce Resources and Priority Ethics: Why Should Maximizers Be More Conservative?," *Ethics Med Public Health*, vol. 18, p. 100698, 2021, doi: 10.1016/j.jemep.2021.100698.

[126] S. A. E. J. P. D. and A. M. Afroogh, "Empathic design in engineering education and practice: An approach for achieving inclusive and effective community resilience," *Sustainability*, vol. 13, no. 7, 2021.

[127] S. Afroogh *et al.*, "Embedded Ethics for Responsible Artificial Intelligence Systems (EE-RAIS) in disaster management: a conceptual model and its deployment," *AI and Ethics*, Jun. 2023, doi: 10.1007/s43681-023-00309-1.

[128] S. A. A. E. M. M. K. and H. A. Afroogh, "Trust in AI: progress, challenges, and future directions," *Humanit Soc Sci Commun*, vol. 11, no. 1, pp. 1–30, 2024.

[129] S. Afroogh, "A probabilistic theory of trust concerning artificial intelligence: can intelligent robots trust humans?," *AI and Ethics*, vol. 3, no. 2, pp. 469–484, 2023.

[130] M. Anderljung *et al.*, "Frontier AI Regulation: Managing Emerging Risks to Public Safety," 2023, *arXiv*. doi: 10.48550/arXiv.2307.03718.

[131] J. Visave, "AI in Emergency Management: Ethical Considerations and Challenges," *Journal of Emergency Management and Disaster Communications*, vol. 05, no. 01, pp. 165–183, Jun. 2024, doi: 10.1142/S268998092450009X.

[132] Deirdre Ahern, "The New Anticipatory Governance Culture for Innovation: Regulatory Foresight, Regulatory Experimentation and Regulatory Learning," *European Business Organization Law Review*, 2025.

[133] "AI Training for Government Officials: Bridging the Knowledge Gap," *resiliencebuilder.co*, Jan. 2024.

[134] J. Chen, V. Storchan, and E. Kurshan, "Beyond Fairness Metrics: Roadblocks and Challenges for Ethical AI in Practice," *arXiv*. doi: 10.48550/arXiv.2108.06217.

[135] R. Soden, D. Wagenaar, D. Luo, and A. Tijssen, "Taking Ethics, Fairness, and Bias Seriously in Machine Learning for Disaster Risk Management," *arXiv*. doi: 10.48550/arXiv.1912.05538.